# Crown-Like Structures in Breast Adipose Tissue: Finding a "Needle-in-a-Haystack" using Artificial Intelligence and Collaborative Active Learning on the Web

Praphulla MS Bhawsar[1], Cody Ramin[1,2], Petra Lenz[3], Máire A Duggan[4], Alexandra R Harris[1], Brittany Jenkins[1,5], Renata Cora[6], Mustapha Abubakar[1], Gretchen Gierach[1], Joel Saltz[7], Jonas S Almeida[1]

1 - Division of Cancer Epidemiology and Genetics, National Cancer Institute, National Institutes of Health, Maryland, USA
2 - Cancer Research Center for Health Equity, Cedars-Sinai Medical Center, Los Angeles, CA, USA
3 - Leidos Biomedical Research Inc, Frederick, Maryland, USA
4 - Department of Pathology and Laboratory Medicine, Cumming School of Medicine, University of Calgary, Alberta, Canada
5 - Department of Biochemistry and Molecular Biology, Johns Hopkins Bloomberg School of Public Health, Baltimore, MD, USA
6 - Stamford, CT, USA
7 - Department of Biomedical Informatics, Stony Brook University, Stony Brook, NY, USA

**Abstract**
Crown-like structures (CLS) in breast adipose tissue are formed as a result of macrophages clustering around necrotic adipocytes in specific patterns. As a histologic marker of local inflammation, CLS could have potential diagnostic utility as a biomarker for breast cancer risk. However, given the scale of whole slide images and the rarity of CLS (a few cells in an entire tissue sample), microscope-based manual identification is a challenge for the pathologist. In this report, we describe an artificial intelligence pipeline to solve this needle-in-a-haystack problem. We developed a zero-cost, zero-footprint web platform to enable remote operation on digital whole slide imaging data directly in the web browser, supporting collaborative annotation of the data by multiple experts. The annotated images then allow for incremental training and fine tuning of deep neural networks via active learning. The platform is reusable and requires no backend or installations, thus ensuring the data remains secure and private under the governance of the end user. Using this platform, we iteratively trained a CLS identification model, evaluating the performance after each round and adding examples to the training data to overcome failure cases. The resulting model, with an AUC of 0.90, shows promise as a first-pass screening tool to detect CLS in breast adipose tissue, considerably reducing the workload of the pathologist.

Platform available at: https://episphere.github.io/path

## 1. Introduction

Composed of dead or dying adipocytes surrounded by CD68+ macrophages, crown-like structures (CLS) are considered a hallmark of adipose tissue inflammation and have been shown to be associated with obesity, an unfavorable metabolic profile, and increased risk of breast cancer [1,2,3]. The etiology of CLS formation in breast adipose tissue has been linked to immunomodulatory molecules, such as cytokines and interleukins, which can impact the proliferation and progression of breast cancer cells [4]. Biopsies of benign breast precursor lesions have been shown to harbor more CLS than normal breast tissue [1,2].

Despite its potential as a biomarker for breast cancer risk, further investigation is needed to better elucidate the role of CLS in breast carcinogenesis.

Methodological issues in CLS quantification, including differences in study design, detection, and pathologist evaluation, have posed challenges in establishing their association with cancer risk [5]. Furthermore, identification of CLS by microscope-based (manual) pathologic review is time-consuming and labor-intensive, posing a significant barrier to conducting the large-scale studies needed in this field. Therefore, there is a clear need for standardization and automation to enable broader study of CLS.

The ability to use Artificial Intelligence (AI) methods to automate the screening of rare features or occurrences enables investigations that would otherwise be unbearably laborious [6,7]. AI-assisted screening could allow even the more extreme cases, such as the one explored in this report, to be handled as an automated or semi-automated module of the clinical workflow.

For AI-based CLS identification in particular, there are two challenges that must be overcome. First, for pooled data such as a whole slide image (WSI), only a few cells typically show CLS. When determining whether a slide contains CLS, a model must first make predictions for each sub-region, also called a patch or tile, at the resolution of individual adipocytes before an aggregated slide-level prediction can be obtained. Most of these predictions will be negative for CLS. The rarity of positive classifications is reflected by the Area under the Receiver Operating Characteristic curve (AUC) of the model approaching the asymptotic 0.5 value for slide-level classification [8]. In other words, the AI model classifying each individual tile will have to make up for that rarity with a high tile-level AUC. This can best be achieved by iteratively improving upon the model via an active learning pipeline, so as to consistently incorporate expert feedback throughout the modeling exercise.

The second challenge is the annotation of the WSI data by pathologists. To start with, whole slide images are typically quite sizeable, with each image being on the order of multiple gigabytes. This makes WSI data hard to operate upon programmatically. Furthermore, as underscored previously, the rarity of CLS in these large images makes their manual identification and annotation a particularly tedious task for pathologists. In the case of this study for instance, each slide, chopped up into 80,000 tiles on average, showed at most one or two instances of CLS, in the already rare instance that the slide contained CLS at all [9]. This is why we describe this problem as akin to finding a needle in a haystack. Such an annotation exercise can best be performed via a software tool where multiple experts can collaboratively view and annotate candidate slides, without compromising on the governance or provenance of the data. The user friendliness of such a tool would also be of critical importance given the vast numbers of tiles to be screened, as would its ease of accessibility across geographical and institutional domains to support seamless collaboration. Finally, a pipeline supporting incremental learning would be optimal, so as to ensure that the model performance can inform the need for further annotation.

To overcome these challenges, we leverage the portability, accessibility and privacy-preserving nature of the web via our digital pathology analytics platform, epiPath [10].

## 2. Methods

### 2.1 Platform
The web platform we developed, called epiPath (see Platform Availability in the Abstract), operates entirely within the context of the user's web browser. The platform connects with the Box.com cloud storage service so that the user merely needs to log into their Box account to access their data on the platform. Note that Box is merely a proxy here for a cloud storage service supporting remote access – the workflow could work the same with data stored on a different service.

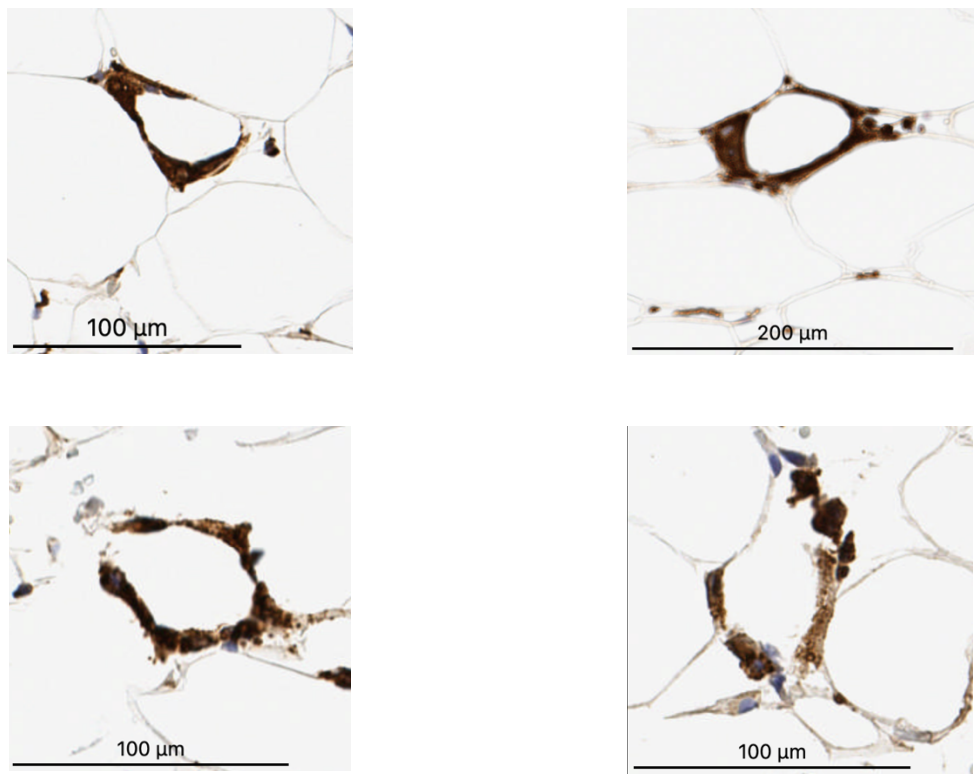

*Fig. 1:* Examples of Crown-Like Structures in Breast Adipose Tissue Whole Slide Imaging data.

In epiPath, the whole slide images, the annotations on them and the trained models are all stored in and operated from Box via their authenticated API, avoiding the complications and potential risks of data downloaded to the local computer or an external server. Since there is no backend, all computation is restricted to the user's own device, and all analytical results are stored back to Box. This ensures that governance is never compromised. Because the file is the single source of truth, this no-downloads approach guarantees robust provenance and traceability at the file level throughout the analysis workflow. Moreover, the shareability of data offered via Box also enables flexible configurations of collaborative arrangements across institutions and geographical locations.

Any roadblocks to accessing the platform are removed because of its client-side operation – there are no costs, no installations and no footprint associated with using it. As a web tool, epiPath is always available via its URL, can be immediately ported to any device with a modern web browser, and, as a software artifact, is completely Findable, Accessible, Interoperable and Reusable (FAIR) [11].

## 2.2 Tile-based access
CLS, in addition to being rare, typically occur in isolation (examples in Fig. 1). Therefore, the AI model never needs to see the entire image. Instead, it can operate on individual tiles of the image at a given point. Once analysis of one tile is finished, the model can move on to the next tile, and in this way, operate on the entire image. This is similar to how a pathologist might look at the slide as well, analyzing a specific region at multiple levels of power without requiring the context of the entire slide when looking for CLS. As a result, the memory and compute requirements are reduced significantly, making it possible to run the workflow even on a commodity computer. We leverage ImageBox3, a web-based tiling mechanism [12], to facilitate remote tile-based operation upon these WSI data. ImageBox3 allows for efficient, on-demand access to tiles of the remote slide from within the browser, never requiring the image file to be downloaded locally. ImageBox3 is integrated within the epiPath platform as the default access mechanism for WSI data. The architecture of the tile-based image analysis pipeline is shown in Fig. 2.

### 2.3 Initial Dataset

Even though CLS form distinct patterns, there is considerable variability in their morphology, depending particularly on whether the macrophages have surrounded the adipocyte completely, also known as a complete CLS (CCLS, adipocyte encirclement > 90%), or partially, known as a borderline CLS (BCLS, adipocyte encirclement 50-90%). It was thus essential to obtain enough examples of CLS for the model to account for this variability and reliably predict tile-level CLS presence. The data for this study was acquired in multiple steps. The initial data acquisition involved 13 WSI that were independently annotated by 3 experts (RC, MA, PL) to contain CLS. The training data was generated from these WSI to contain 224μm x 224μm tiles sampled at a resolution of 512x512 pixels. The tile-level labels for CLS presence or absence were applied via the Annotations functionality within epiPath. Because negative examples significantly outnumbered positive ones, the dataset was augmented with trivial affine transformations applied to CLS-positive tiles in order to create a balanced training dataset. To improve model generalizability, tiles with similar staining as CLS, as well as areas showing necrosis and stroma, were manually selected and added to the training dataset. In all, the initial dataset contained 303 PNG tiles, 106 of which were positive for CLS (both borderline and complete, after dataset augmentation), and 197 with no CLS.

### 2.4 Modeling & Active Learning

A convolutional neural network was trained using Google's AutoML service on these images directly to provide a binary response with respect to CLS presence. The images were not preprocessed in any way to ensure that the model remained portable and could be used outside of the platform as well, where the exact preprocessing method might not be available. After training, the model was exported as a TensorFlow.js package and placed back in Box under the same governance as the original images. It was then run within the epiPath platform to obtain tile-level predictions on a validation set containing whole slide images from a different dataset. A tile would be marked CLS Present if the prediction was positive with a score greater than a certain threshold, which could be tuned from the platform. Based on the training metrics, a threshold of 0.7 was agreed to be optimal in terms of minimizing false positives.

Two reviewers (MA, PL) evaluated the model performance on this dataset within epiPath, either affirming or rejecting the model's predictions. Since these images had not previously been seen by the pathologists, their feedback effectively served as tile-level annotations. Fig. 3 shows the platform facilitating this evaluation by providing an easy interface to agree or disagree with the model's predictions.

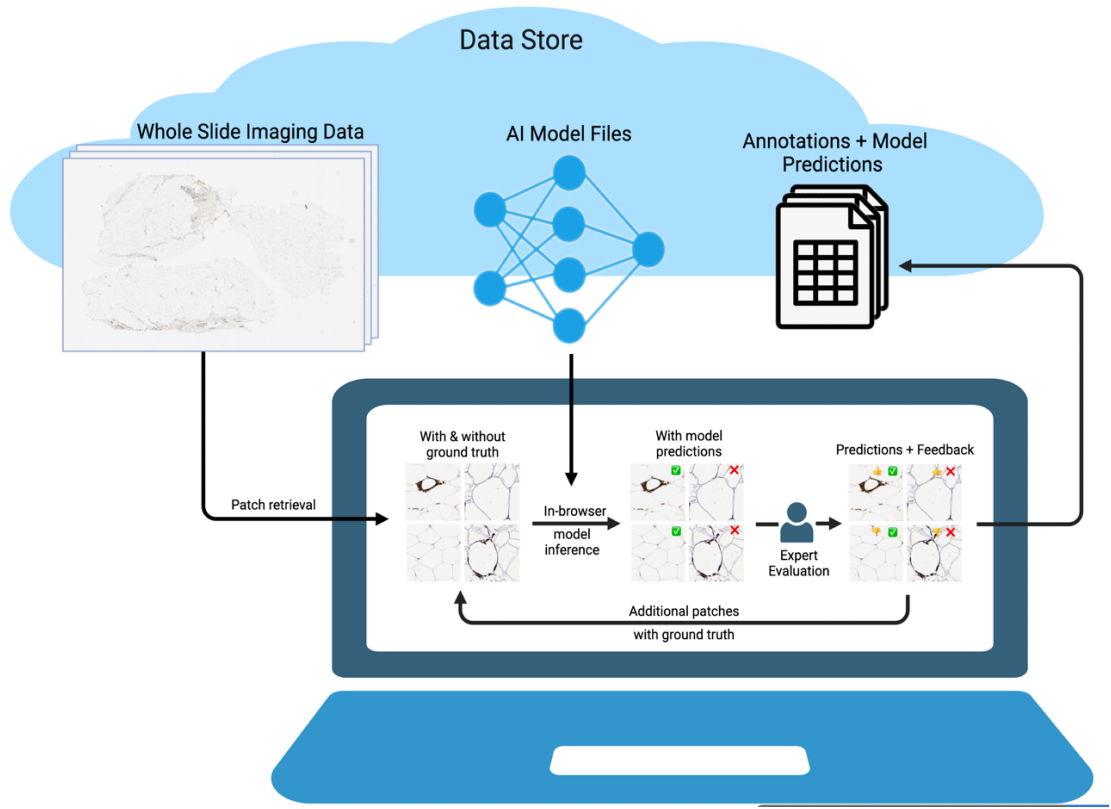

***Fig. 2*** *The tissue-to-analysis pipeline using the epiPath platform for identifying CLS presence. The data store used here was Box.com, housing the WSI data, the model files, predictions and pathologist feedback. Once the clinician/user logs into their Box account via the platform, they automatically have access to their data within the tool, along with any trained models. Only patches of the whole slide image are retrieved, facilitated by ImageBox3. Model inference is computed on these patches, after which the user can evaluate the predictions. The feedback thus provided can then be used as part of the training data for the next iteration of the model.*

Having identified the model's failures, the model predictions along with the reviewers' feedback were saved back to Box and used to train the model from scratch. This new training dataset contained all examples from the previous training set along with the newly identified failure cases from the validation images. A seamless active learning pipeline was thus realized, allowing the model to be trained iteratively and incrementally. With another 302 tiles added to the training dataset (106 CLS positive and 196 negative), the model was retrained and applied to a new validation set. In all, six such rounds of active learning were conducted. The final model was validated on a dataset containing 48 whole slide images, totaling over 3,000,000 tiles. Two pathologists (PL and MAD) independently reviewed the model predictions via the platform. Inter-observer agreement was calculated, and the disagreements were independently reviewed by a third pathologist (MA). The gold standard of CLS presence was then established based on agreement between any two of the three pathologists. This standard was then used to evaluate the performance of the final model.

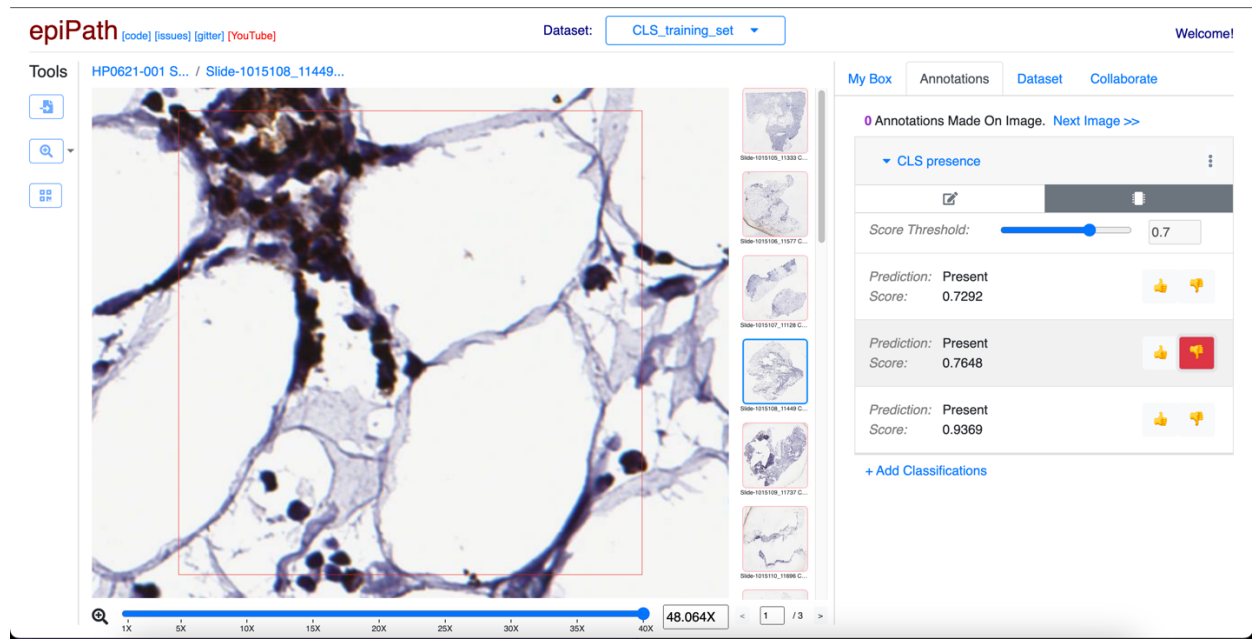

***Fig. 3*** *The epiPath platform being used to evaluate the model's predictions. For every tile predicted upon by the model, the pathologist has the option to affirm or reject the prediction. The selected prediction on the right corresponds to the tile outlined in red within the viewer. The score at which to label a tile as CLS present can be changed using the Score Threshold slider to optimize for sensitivity and/or specificity. The thumbs up/down buttons against the prediction make it easy to provide feedback and serve as ground truth for subsequent active learning iterations.*

## 3. Results

Due to the needle-in-a-haystack nature of this problem as described above, training an AI model that was both sensitive to CLS presence and specific in its predictions would have been difficult to achieve. Moreover, differences in staining across the datasets used and the presence of artifacts were also challenging to handle. The intermediate models trained during the active learning workflow showed reasonably good performance, but they were invariably plagued with false positive results (type 1 errors). Since the objective of this exercise was to validate epiPath as a tool that could reduce the time and effort required by a pathologist to screen for CLS, we focused primarily on reducing the number of false positives through each active learning iteration while ensuring zero false negatives. The final model performed remarkably well at tile-level classification, demonstrating an Area Under the Curve (AUC) of 0.90.

Slide-level predictions were also generated from the tile-level classifications, since using the CLS as a clinically relevant biomarker would require identification by case or subject. A slide was marked as positive for CLS if it contained one or more tiles marked as CLS present by the model. The slide-level ground truth was obtained from the majority-based gold standard as described before. The AUC for this slide-level analysis is close to 0.5. This is a result reflecting the presence of type 1 errors mentioned above. Most of the slides themselves contained no CLS whatsoever, but each slide had on average 80,000 tiles that were evaluated by the model for CLS presence. This made it highly probable that the model would make at least one positive prediction on a given slide. Of the 48 whole slide images used for validation, only 9 were identified to contain CLS by the pathologists. In all, there were only 14 tiles with CLS across all 48 WSI, an occurrence rate of 0.0004%. It is difficult to reduce the false positive rate to this extent without overfitting the model. Therefore, this aggregate analysis is not entirely reflective of the model's performance. Fig. 4 shows the Receiver Operating Characteristic (ROC) curves for the final model.

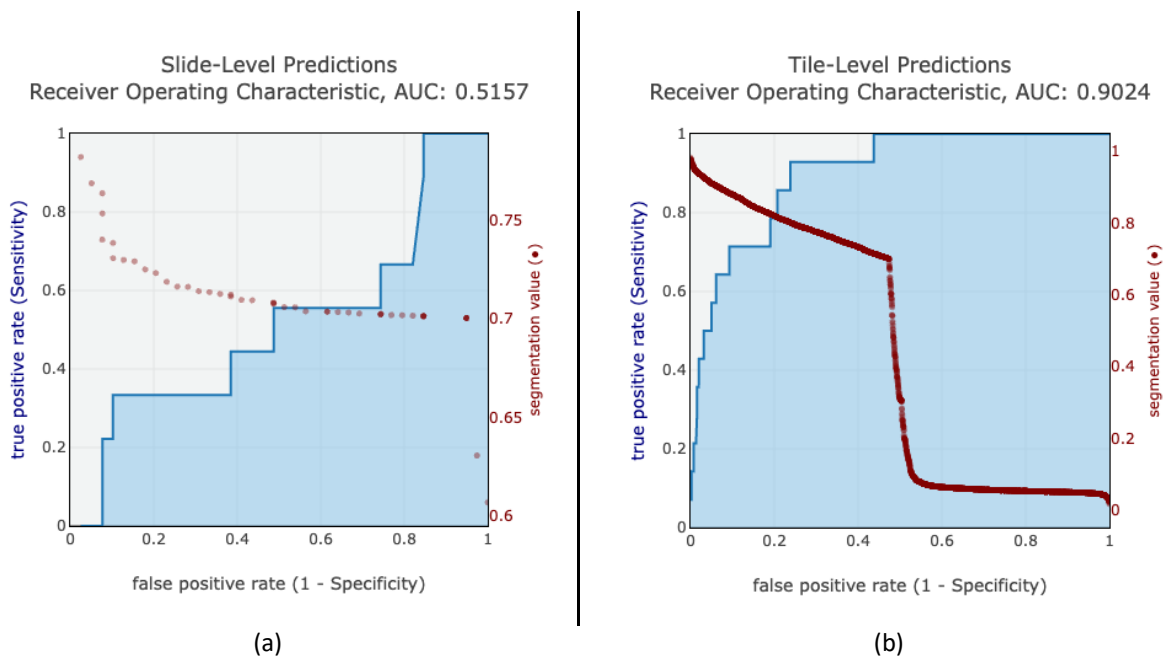

*Fig. 4:* The Receiver Operating Characteristic (ROC) curves of the (a) slide-level, and (b) tile-level predictions for CLS presence by the final model. The significant difference between the AUC scores reflects the fact that each slide has a large number of tiles to be classified, and the slide-level predictions aggregate the tile-level scores such that a single positive tile causes the slide to be labeled CLS positive. The effective tile-level classification enables the automation of what would otherwise be an extremely tedious process: that of finding rare regions of interest (ROI) that could then be manually evaluated by an expert. Note that the choice of 0.7 as the score threshold for positive classification, along with the derivation of the ground truth from the model predictions using that threshold, leads to the sharp drop at that score in the tile-level ROC segmentation plot.

## 4. Discussion

The approach taken in this study is, to the best of our knowledge, the first example of a completely in-browser AI annotation-to-inference pipeline for whole slide imaging data. While the solution has its challenges, most notably having to operate within the limited compute and memory resources of the user's machine, the benefits of user-centric governance, cost-effectiveness (no nominal costs), FAIR-ness [13], and portability far outweigh them. This is most striking in the context of the reproducibility crisis in biomedical AI research [14]. In addition, separating the analytics environment from the data makes collaboration much easier. Access to the environment by a collaborator does not automatically imply access to the data. Instead, access is resolved directly at the API level (each user sees the data they have permission for), considerably simplifying governance. In essence, the code travels to the data, instead of the data having to be made locally available to the code. Furthermore, security and ease of delivery come as a side-effect of using the web browser as the software execution environment, allowing for the analytics to take prime focus. We believe that the results in this study validate the potential of the web as a means to distribute analytical artifacts for use by the community.

Tile-level analysis of the image, while necessary when operating on large WSI data, comes with a few complications. A CLS on the slide might, for instance, span multiple tiles. While this is a rare event (e.g., of the 14 CLS identified by the model in the validation set, only one spanned multiple tiles), it may lead to spurious results. Since the model focuses on only a single tile at a time, it may make multiple positive predictions or misses on a few tiles containing parts of the CLS. In this study, CLS spanning multiple tiles were recorded only once by the pathologists and assigned to the tile which contained most of the CLS. Alternatively, this could be resolved with a downstream method that aggregates adjacent positive tiles into composite tiles that are marked for CLS presence. The configuration of that post-processing step can be followed by image segmentation to highlight the exact location of the predicted CLS. The model also does not distinguish between complete and borderline CLS, and this could be histologically relevant. Since identification at the tile-level is reliably achieved with our proposed model, the measurement of the adipocyte encirclement becomes possible as a downstream task in the same or different computational environments.

In this work, we approached the modelling as a means to develop a preliminary screening method to identify a rare morphological structure, CLS, from the comparatively vast sea of breast adipose WSI data. This is a scenario where the model is precariously susceptible to false positives. Accordingly, the task of increasing model specificity is balanced against the main analytical goal of this study – to reduce the number of tiles that a pathologist would have to survey by at least two orders of magnitude. For the CLS modelling exercise described in this report, a slide that typically required hours can be analyzed in minutes, by only reviewing the tiles screened by the model. The very low false negative rate reflects the usefulness of the model as a preliminary screening test for CLS, and potentially for breast cancer development and progression.

**Conclusion**
The traditionally siloed nature of WSI analytics creates obstacles to reproducibility and collaborative analytics. However, with the advent of efficient web computing, it is now possible to break the proverbial silos open and run data-intensive analytical workflows on the web browser – an openly available execution environment that addresses security, privacy, and availability at no cost to either the clinician or the developer of the analytical tool. We prove the viability of this approach for challenging problems similar to Crown-Like Structure identification in this report. Given the ease of collaborative annotation and active learning, the availability of portable image analytics as a web-accessible, free to use software module could be highly impactful for histopathology research.